\documentclass[acmsmall]{acmart}
\acmJournal{PACMHCI}
\setcopyright{usgov}
\acmYear{2024} \acmVolume{8} \acmNumber{CSCW2} \acmArticle{415} \acmMonth{11} \acmDOI{10.1145/3686954}
\usepackage{amsmath,array,multirow,graphicx,pdflscape}
\usepackage{bm} %
\usepackage[utf8]{inputenc}
\usepackage[flushleft]{threeparttable}

\AtBeginDocument{%
  \providecommand\BibTeX{{%
    \normalfont B\kern-0.5em{\scshape i\kern-0.25em b}\kern-0.8em\TeX}}}

\setcopyright{usgov}
\acmYear{2024} 
\acmVolume{8} 
\acmNumber{CSCW2} 
\acmArticle{415} 
\acmMonth{11}
\acmDOI{10.1145/3686954}

\usepackage{graphicx} %
\usepackage{todonotes}
\usepackage[inline,nomargin,index]{fixme}
\usepackage{url}

\usepackage{natbib}

\def\citepos#1{\citeauthor{#1}'s \cite{#1}}

\fxsetup{status=final,theme=color,mode=multiuser,inlineface=\itshape,envface=\itshape}

\author{Chau Tran}
\affiliation{%
  \institution{New York University}
  \city{New York, NY}
  \country{United States}
}
\email{chau.tran@nyu.edu}

\author{Kejsi Take}
\affiliation{%
  \institution{New York University}
  \city{New York, NY}
  \country{United States}
}
\email{kejsitake@nyu.edu}

\author{Kaylea Champion}
\affiliation{%
  \institution{University of Washington}
  \city{Seattle, WA}
  \country{United States}
}
\email{kaylea@uw.edu}

\author{Benjamin Mako Hill}
\affiliation{%
  \institution{University of Washington}
  \city{Seattle, WA}
  \country{United States}
}
\email{makohill@uw.edu}

\author{Rachel Greenstadt}
\affiliation{%
  \institution{New York University}
  \city{New York, NY}
  \country{United States}
}
\email{greenstadt@nyu.edu}

\FXRegisterAuthor{ct}{act}{\color{purple}Chau}
\FXRegisterAuthor{kc}{akc}{\color{blue}Kaylea}
\FXRegisterAuthor{rg}{arg}{\color{red}Rachel}
\FXRegisterAuthor{mh}{amh}{\color{green}Mako}
\FXRegisterAuthor{kt}{akt}{\color{orange}Kejsi}

\title{Challenges in restructuring community-based moderation}
\date{February 2023}

\begin{document}

\begin{abstract}
    Content moderation practices and technologies need to change over time as requirements and community expectations shift. However, attempts to restructure the existing moderation practices can be difficult, especially for platforms that rely on their communities to moderate, because changes can transform the workflow and workload participants' reward systems. By examining the extensive archival discussions around a prepublication moderation technology on Wikipedia named \textit{Flagged Revisions}, complemented by seven semi-structured interviews, we identify various challenges in restructuring community-based moderation practices. Thus, we find that while a new system might sound good in theory and perform well in terms of quantitative metrics, it may conflict with existing social norms. Furthermore, our findings underscore how the relationship between platforms and self-governed communities can hinder the ability to assess the performance of any new system and introduce considerable costs related to maintaining, overhauling, or scrapping any piece of infrastructure.
\end{abstract}

\maketitle

\section{Introduction}

Majority of all online platforms employ content moderation in some way. Content moderation is often necessary to meet community needs, maintain the reputation and integrity of platforms, comply with legal requirements, and create an advertisement-friendly environment \cite{10.1145/3534929, grimmelmann2015virtues}. Over time, we expect content moderation practices to evolve and adapt as community expectations shift, unwanted behaviors manifest, new ideas emerge about effective moderation practices, and legal frameworks are revised. Hence, online platforms often review and adjust their moderation practices by introducing new guidelines, approaches, and technologies~\cite{kalsnes2021hiding, caplan2018dead}.

We present a case study of \textit{Flagged Revisions} (also known as \textit{FlaggedRevs}), a prepublication moderation system \cite{7975786} on Wikipedia, as well as its subsequent scaled-back deployment under the name \textit{Pending Changes}. FlaggedRevs was designed to reduce the impact, and perhaps even the prevalence, of vandalism on wikis. The system achieved early success and was adopted by dozens of wikis run by the Wikimedia Foundation (WMF) \citep{chau_tran_flaggedrevs}. The founder of Wikipedia, Jimmy Wales, was among its most vocal supporters and actively called for it to be installed on every WMF wiki including Wikipedia's more than 300 different language editions. 
However, as more communities adopted it, users expressed concerns. Despite attempts to address these concerns, community support for FlaggedRevs dwindled over time and WMF halted its further deployment indefinitely. Eventually, they placed its code into ``maintenance-only'' mode. 

FlaggedRevs reflects an empirical puzzle for social computing researchers, designers, and practitioners. The Wikimedia community designed the system, and WMF leadership supported its adoption. Further, from a quantitative perspective, it seems to have been a resounding success at achieving its stated design goals. \citet{chau_tran_flaggedrevs} found that the system substantially reduced the amount of vandalism visible on the site with little evidence of negative trade-offs. 

Although puzzling, FlaggedRevs' story is far from unique. In a classic work in social computing, Orlikowski and Gash showed that new technology is often met with unexpected resistance caused by conflicts with users' mental models and structural features of communities \citep{orlikowski1991changing, orlikowski92fromnotes, orlikowski1994technological}. Perhaps FlaggedRevs' downfall is a consequence of relationships with the system's stakeholders?
Of course, \citeauthor{orlikowski1991changing}'s work is based on very different technologies (desktop software applications) in very different settings (traditional firms).
Our work extends the scholarship by \citeauthor{orlikowski1991changing} by returning to a version of their core question but focused on the type of content moderation system in the types of organizations that lie at the core of contemporary social computing scholarship: \textit{Why might a community-based moderation system fail to be widely adopted?} 

To answer this question, we conducted an extensive archival analysis of 67 documents related to implementing and discussing FlaggedRevs on different Wikipedia language editions and seven semi-structured interviews with Wikipedia administrators and moderators. 
We organize our observations about challenges along three dimensions, namely, \textit{community}, \textit{platform and policies}, and \textit{technology}. Overall, our findings suggest that while a new technology might demonstrate positive effects, it may still conflict with the platform's social structure in the form of policies, norms, and reward systems.
While extensive research in social computing has studied attempts to integrate new technologies in different organizational settings \citep[e.g.,][]{kiene2019technological, leonard2011implementation, barrett2012reconfiguring}, our studyfocuses on how this process plays out in a community-oriented platform with a high degree of self-governance. Our work also makes an empirical contribution by suggesting a solution to the puzzle raised in \citepos{chau_tran_flaggedrevs} quantitative study of FlaggedRevs (i.e., why this particular system with strong performance metrics might ultimately still fail). Furthermore, we make a theoretical contribution by placing the older social computing literature on technological change in organizations in direct conversation with the more contemporary literature on content moderation and shared governance. Our case study also shows how this older literature on technological change in organizations can be updated and translated in ways that allow it to continue to be relevant to contemporary social computing contexts.
The findings also suggest that the complex relationship between online platforms that rely on community-based moderation, such as Reddit, Discord, and Twitch, and their self-governed communities can prevent both sides from providing necessary support for the technology they both hope to adopt.

\section{Background}
\label{sec:background}
Our analysis of the case of FlaggedRevs draws from three lines of prior research in social computing literature, namely, peer production, moderation and governance, and technological change. To provide context for our study, we introduce each of them.

\subsection{Peer production and Wikipedia}
Peer production is a form of internet-mediated collaborative practice focused on open creation and sharing, performed by online groups that set and execute goals in a decentralized manner \cite{benkler2015peer}. The concept of peer production was initially introduced by Benkler in 2002 who identified two evolutionary phases of successful peer production communities, namely, (1) creating content (``utterance'') and (2) quality control (``relevance/accreditation'') \cite{benkler2002coase}.
Often, these two goals are interconnected. For example, social computing research on Wikipedia argued that the goal of maintaining contributors is a key feature of the community's success \cite{halfaker2013rise, halfaker11dontbite, bryant2005becoming, suh2009singularity}. 

However, as Wikipedia has grown, it erected barriers to entry to discourage misbehavior %
\cite{halfaker2013rise}. This pattern appears to be a general feature in large peer production communities \citep{teblunthuis_revisiting_2018}. For example, citing concerns around low-quality content and abuse, peer production communities, including a growing number of Wikipedia communities, instituted new requirements for users to create accounts to contribute \citep{hill2021hidden, wiki_unregistered_editing}.  Research shows that doing so reduces both high and low-quality contributions from registered and unregistered participants \cite{hill2021hidden}. Similarly, hostile responses to low-quality contributions from newcomers can discourage newcomers' continued participation \cite{halfaker11dontbite, halfaker2013rise}. This study points to a key tradeoff in the design of production systems between participation and content quality. When designed well, a barrier to entry can act as a filter to increase average contribution quality \citep{kraut2012building}. However, these barriers are never perfect filters and typically increase the cost of participation \citep{kraut2012building, hill2021hidden}.
An example of this tradeoff in social computing literature relates to the use of automated tools in quality control in Wikipedia. \citet{halfaker2013rise} present evidence that the use of these tools negatively affects the experience of new contributors and leads to a decrease in the number of active editors on English Wikipedia.

\subsection{Community-based moderation}

For over two decades, social computing research on content moderation in peer production settings---Slashdot, Wikipedia, Reddit, and more---has focused on the social dynamics that drive community-based forms of moderation \cite{gillespie2018custodians, kiene19volunteer, seering_reconsidering_self_moderation, lampe2004slash, poor2005mechanisms}.  Seering defines ``community-based moderation'' as a form of moderation performed mainly by community members, with minimum direct interference from platform administrators \cite{seering_reconsidering_self_moderation}, in contrast to top-down and platform-centric content moderation models used on large corporate social media platforms such as YouTube, TikTok, Twitter/X, or Facebook \cite{seering_reconsidering_self_moderation, forte2009decentralization}.

In community-based moderation, members act as both contributors and moderators to shape and maintain local norms \cite{matias2019civic, kraut2012building, gillespie2018custodians}. These community-based moderators are understood to be acting out of a passion for their communities, typically sharing common interests and ideologies, and working as volunteers \cite{bryant2005becoming, ostrom1990governing, seering2019moderator}. The efforts of community-based moderators are invaluable for peer production platforms such as Wikipedia, Reddit, and Discord \cite{seering_reconsidering_self_moderation}.

If \textit{content moderation} describes the action of removing content and banning users, \textit{governance} describes the process through which the rules that direct these actions are devised. In Wikipedia, like most other community-based moderation sites, content governance is also decentralized. Community-based moderation often occurs through with what \citet{ostrom1990governing} describes as ``polycentric'' governance, in that rule-making occurs on multiple levels. Although community members receive considerable freedom and power in developing, deploying, and enforcing moderation policies and tools, they typically must adhere to platform-level rules as well \cite{seering_reconsidering_self_moderation, de2012coercion, thach2022visible}.
In this sense, governance on Wikipedia is similar to other peer production communities \cite{jhaver2023decentralizing}.
For example, Reddit contains millions of subreddit communities and StackExchange contains approximately two hundred Q\&A websites.  %
In an example of polycentric governance, Reddit presents a \textit{Moderator Code of Conduct} page that lists general rules for all moderators on the platform while also noting that moderators ``are at the frontlines using [their] creativity, decision-making, and passion to create fun and engaging spaces for [other members]'' \cite{reddit_coc}. As long as moderators adhere to Reddit's four basic rules, they are free to moderate content as they see fit.

On platforms governed following a polycentric model, users may actively participate in the creation and refinement of moderation policies~\cite{forte2009decentralization}. Although this gives users a more active role in shaping and maintaining social norms, democratic governance can also hinder efficient decision-making and long-term planning \cite{de2012coercion}. Furthermore, while platform-level governing bodies may exhibit a hands-off approach regarding content moderation, they are often relied upon to provide configurable moderation technologies to assist users across different communities who seek to moderate content differently. For example, Reddit developed a highly configurable program called \textit{AutoModerator} that can be used in all subreddits \cite{konieczny2009governance}. 

Similarly, WMF developed tools such as the \textit{Objective Revision Evaluation Service (ORES)}, which uses machine learning to assess the probability that the revision will be damaging \citep{halfaker_ores_2020} and \textit{Edit Filters}, which applies basic heuristics about unwanted contributions to prevent certain revisions, warn users against making them, or tag edits for review \citep{vaseva_you_2020}. Furthermore, automated moderation tools can prove helpful for community-based moderation when communities, reliant on volunteers, demonstrate insufficient resources for moderation activities \cite{gillespie2020content}. To address these labor shortages, many communities develop their own moderation tools including bots \cite{geiger_bots_2014, kiene20bots} and external applications \citep{geiger_work_2010}.

\subsection{Technological change in organizations}

A large body of work in social computing, management, and organizational studies invesgiates better ways for organizations to manage and navigate technological change \cite{allen1984managing, oliver2011technological, orlikowski1996improvising, kiene2019technological}.  An important stream of this work in social computing suggests that technological change can be disruptive to communities and can result in unintended consequences \cite{kiene2019technological, leonard2011implementation, barrett2012reconfiguring, orlikowski1991changing, orlikowski92fromnotes, orlikowski1994technological}. These unintended consequences can fundamentally shape social relations and can arise due to either a lack of foresight and proper planning or the sheer unpredictability of complex sociotechnical processes \cite{dafoe2015technological}. 

Orlikowski and Gash \citep{orlikowski1991changing, orlikowski1994technological, orlikowski92fromnotes} present evidence from a rich ethnographic study of the deployment of Lotus Notes in a large consulting firm. The deployment, the authors argue, was largely unsuccessful in accomplishing the goal of transforming work in the firm. Orlikowski and Gash argue that part of the reason for this failure was that the changes demonstrated different and unintended impacts across groups within the organization. One important set of challenges relates to \textit{technological frames}---a term coined by Orlikowski and Gash to describe the shared understanding and knowledge needed to use a tool effectively. 
When users' mental models of the tool are incompatible with its intended function, they will have trouble adopting the tool effectively. Their frame of reference becomes what Orlikowski and Gash call ``incongruent'' \cite{orlikowski1991changing}. Subsequently, Orlikowski and Gash showed that when the level of technological change is substantial (i.e., more than an incremental update), users can struggle to adapt to the change due to their incongruent technological frame.

Although Orlikowski and Gash's work laid important groundwork and was deeply influential in early CSCW scholarship in various settings, their work was focused on firms and organizations with centralized governance models. %
We join \citet{kiene2019technological} in exploring how these classic perspectives can be applied to community-based moderation in peer production communities. While \citeauthor{kiene2019technological} explores the role of user innovation in deploying new moderation solutions, we conduct a study that is similar, in some respects, to Orlikowski and Gash's classic study on Lotus Notes---but in the context of community-based moderation and community-based governance. In this different context, we seek to understand to what degree their theory applies and how it might be updated to reflect the specific challenges of deploying social computing tools in these new, less formal settings. %

\section{Empirical Setting}
\label{sec:empirical}

Wikipedia hosts encyclopedia-building communities working in more than 300 different languages, often referred to as \textit{language editions}, exhibiting vastly different sizes and user bases. Each language edition is semi-autonomous and decides many of its own processes around what content is and is not allowed \cite{jhaver2023decentralizing}. Within Wikipedia, moderation tools may be deployed at different \textit{levels}. Tools may exist at the \textit{wiki level} if they are deployed at the level of an entire language edition. As these tools affect everything in a given language edition, wiki-level tools must be integrated into the website's server-side codebase or configuration. In practice, any significant change to wiki-level tools must be approved and conducted by the WMF.

Most tools, however, are deployed at the \textit{user level}. These tools are often operated by users, perhaps with administrator rights, and are not integrated into the server codebase. Moreover, they are often run at a scale that is smaller than the entire language edition, but it is not required. For the purpose of automated moderation, these tools are typically \textit{bots} \cite{automated_moderation}. These user-level tools are a form of ``bespoke code,'' a term coined by \citet{geiger_bots_2014}. Although it may depend on the task that the bot aims to automate, bot operators typically do not need WMF's approval to deploy a bot. In many cases, however, rules designed by each language edition govern the deployment and usage of bots \cite{bot_policy}.

\begin{figure}
 \centering
  \includegraphics[width=\textwidth]{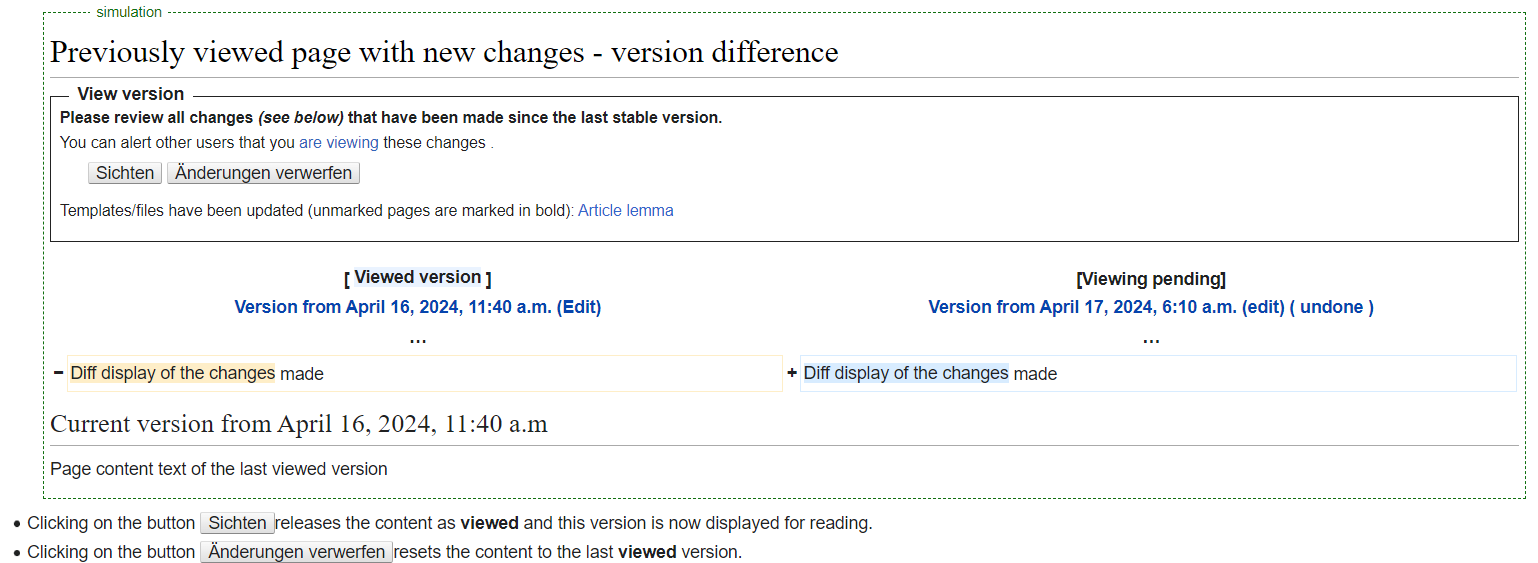}
  \caption{An example of reviewing an pending edit by comparing it to the last``viewed'' edit, which is considered the stable version. This is a screenshot from \cite{viewed_version}, which is translated from German to English using Google Translate, hence it refers to the act of flagging or reviewing an edit as ``viewing''.}
 \label{fig:flaggedrevs_screenshot}
\end{figure}

\textit{Flagged Revisions} (also known as FlaggedRevs) is a wiki-level tool. It was created by developers and contractors from \textit{Wikimedia Deutschland} (an independent nonprofit organization founded to support Wikimedia projects in Germany \cite{wikimedia_deutschland}), with the goal of ensuring that Wikipedia readers only see high-quality versions of articles. German Wikipedia was the first language edition to deploy the system in 2008 \cite{flaggedrevs_enabled}. As the name suggests, FlaggedRevs places a ``flag'' on contributions reviewed by trusted users to indicate that those changes have been reviewed and/or passed a quality check. 
This ``prepublication'' approach to moderation ensures that no edits made by untrusted users are published, at least by default, without first being screened and approved by a trusted user. As shown in Figure \ref{fig:flaggedrevs_screenshot}, users with reviewer rights can review a pending unreviewed edit and decide whether to approve it by marking it as \textit{``viewed''}. They can discard the changes by ``reverting'' them. 

FlaggedRevs was thought to be useful not only because it reduces the visibility and impact of vandalism and substandard contributions but also because some believed it would reduce the incentive to make these types of damaging edits in the first place. After one year of using FlaggedRevs on a trial basis, German Wikipedians voted to continue using the system, with 905 members voting in favor, 362 members voting against, and 33 members formally abstaining \cite{german_wiki_poll}. Furthermore, FlaggedRevs was also enthusiastically supported by Wikipedia founder Jimmy Wales who said, ``I've been pushing for this and waiting for this for years now'' \cite{fr_petition}. He expressed his support for the idea of deploying the tool across all language editions on Wikipedia. Over time, FlaggedRevs was deployed to 25 Wikipedia language editions \cite{flaggedrevs_main} as well as a range of other projects run by the WMF, including Wiktionary, Wikibooks, Wikinews. 

The FlaggedRevs software is designed to be flexible. Individual language editions can adjust an enormous number of parameters that control things such as whose contributions will be subjected to review (i.e., which revisions will be flagged or unflagged by default), minimum thresholds to qualify for reviewer status and the ability to apply flags, and whether and how users are automatically promoted by the system into groups who whose work is unflagged or automatically flagged. Initially, most wikis either chose to adopt German Wikipedia's configuration of FlaggedRevs or relatively small adjustments to it.

WMF's and the Wikipedia community's enthusiasm for FlaggedRevs waned enormously over time. After numerous trial runs and configuration tweaks, many Wikipedia editors continued to express concerns about the potential impact of FlaggedRevs on the health and growth of their communities, leading to several heated debates that eventually caused the WMF to indefinitely suspend FlaggedRevs deployment on new Wikis in 2017 \cite{fr_norwegian}. Recent research published in CSCW showed that many of these concerns, especially claims that it would result in decreased editor activity and new user retention, seem to demonstrate empirical support that is weak at best \cite{chau_tran_flaggedrevs}. In quantitative terms, FlaggedRevs appears to be incredibly effective at keeping low-quality content out of the public eye with little evidence of any large effect (good or bad) on other measures of community health \cite{chau_tran_flaggedrevs}. This reflects an important empirical puzzle that maps to the specific version of the conceptual research question presented earlier: \textit{What led to the decline in FlaggedRevs' popularity over the years, despite its effectiveness?} FlaggedRevs' uneven rise and decline in popularity across language editions presented us with an interesting case study that highlights the important dynamics around unsuccessful attempts at technological change in social computing systems, both in general and in the specific case of content moderation.

\section{Methods}
\label{sec:methods}
Our methods involve a detailed case study in two parts. First, we conduct a qualitative forensic analysis of detailed archival data of discussions. Subsequently, we triangulate findings from this analysis using data from a series of seven interviews and analyze our data using the constant comparative method per \citet{charmaz2014constructing}.

\subsection{Qualitative Forensic Analysis}

Our textual data collection process was \textit{qualitative forensic analysis}, a method described by \citet{champion2019forensic}. Qualitative forensic analysis (QFA) is a method involving the meticulous collection and examination of documents using techniques that bear some similarity to traditional archival methods or investigative journalism. \citeauthor{champion2019forensic} conducted their research in Wikipedia, suggesting that QFA might be particularly appropriate for our setting.
Although they were published before \citet{champion2019forensic} and did not use the term QFA, several examples of other Wikipedia research have used similar data collection and analysis methods \citep[e.g.,][]{10.1145/2145204.2145266, twyman17blm}.

\citeauthor{champion2019forensic} argues that QFA attempts to mitigate the absence of physical participants to help researchers ``reconstruct the participant’s experience'' \cite{champion2019forensic}. Furthermore, QFA seeks to give researchers an understanding of the measures, tools, and strategies employed by subjects who seek to gain an understanding of the phenomenon themselves. Finally, they argue the technique is particularly appropriate for collecting context-rich evidence from participants who are not easily identifiable or reachable, but who demonstrate intimate knowledge of the research subject. %
 
Our textual data collection comprises 62 pages from the following Wikipedia language editions, namely, English, German, Indonesian, and Hungarian. These documents were gathered from four different websites run by the WMF: \textit{Wikipedia}, \textit{MediaWiki}, the request-tracking system \textit{Phabricator}, and \textit{Meta}. Meta, which we refer to as \textit{Meta-Wiki} to avoid confusion with other websites, is used by the Wikimedia community for various purposes ranging from coordination and documentation to planning and analysis. 

Within language-specific Wikipedia editions and Meta-Wiki, our corpus is drawn from several different areas on each wiki called ``namespaces.'' The software used to run these wikis (\textit{MediaWiki}) uses the concept of namespaces to separate different types of pages. Namespaces are indicated in the page name itself by the inclusion of reserved words that are recognized by the MediaWiki software. In our case, these include the \textit{Talk}, \textit{Project}, \textit{Project Talk}, and \textit{Special} namespaces.
Wikipedia articles, for example, are in the ``Article'' namespace, indicated by the absence of a reserved word. Discussion pages are in the ``Talk'' namespace, indicated by prefixing the article name with ``Talk'':.
For example, the encyclopedia article about CSCW is simply called ``Computer-supported cooperative work'', while the discussion page about the article is prefixed with ``Talk:''. The ``Project'' namespace is typically named with the name of the project itself.
One Project page of interest is a wiki-wide discussion forum called the ``Village Pump'' on English Wikipedia (similar forums under different names exist in other language communities). The ``Special'' namespace includes ``pages generated by the software on demand for special purposes, usually related to project maintenance.'' %

From each of the namespaces aforementioned, we sought to collect all material related to the implementation and deployment of FlaggedRevs. The first two authors began data collection by querying keywords that refer to the moderation system. Although the system is called ``Flagged Revisions'' in English, its name varies within English and across languages. For example, on English Wikipedia, we used the keywords: \textit{flagged revisions}, \textit{flaggedrevs}, \textit{pending changes}, \textit{deferred changes}, \textit{patrolled revisions}, \textit{timed flagged revisions}, and \textit{stable pages}. Some of these keywords describe failed proposals to reconfigure and rebrand FlaggedRevs.

Due to the language limitations of our team, we could not search for documents related to the system in every language edition that used FlaggedRevs with equal effectiveness. However, we made an effort to do so in a number of non-English language editions. Since German Wikipedia was the first language edition to enable FlaggedRevs, we identified the local name of Flagged Revisions in German, where it is referred to as \textit{gesichteten versionen}. The first two authors used these keywords to query both the Wikipedia search bar and Google. Moreover, as we cannot fluently read documents in all of the languages in which FlaggedRevs was enabled, we relied on Google Translate to obtain a rough translation of pages and only included those that could be translated and understood by our team. In addition to German Wikipedia, we included translated documents from Indonesian, Farsi, and Hungarian Wikipedia, often based on recommendations of members of these communities that we interacted with as part of our interview study.

Our initial corpus contained thousands of documents but many were not relevant. We excluded all duplicated pages, and manually examined all pages to determine if their content met the following inclusion criteria:

\begin{itemize}
    \item \textbf{Discussion/Talk/Village pump pages}: Did the page contain discussions and consensus polls among users about the deployment of the software or related policies?
    \item \textbf{Meta-Wiki and project namespace pages}: Did the page describe either a proposal related to FlaggedRevs or some discussion around its functionality or deployment?
    \item \textbf{Special pages}: Did the page report the deployment status of the system or provide relevant statistics regarding its performance?
    \item \textbf{Phabricator pages}: Did the task page propose a task to add a feature to FlaggedRevs or reconfigure the software (i.e., we excluded pages about minor bug fixing)?
\end{itemize}

We followed reference links on each page and traced the contribution history of prominent Wikipedia administrators who were directly involved with the deployment of FlaggedRevs to gather additional pertinent documents. Subsequently, we collected FlaggedRevs statistics regarding the time edits spent before being reviewed. These data are summarized and reported in Table \ref{tab:waittime} in our Findings section. 

After screening, our dataset included 49 Discussion/Talk/Village pump pages, 16 Meta-Wiki/Wikipedia namespace pages, 35 Special namespace pages, and 62 Phabricator pages. Unfortunately, the detailed use of technical jargon limited the effectiveness of machine translation tools and meant we had only a limited understanding of some documents not written in English. However, we were able to collect 23 additional discussion pages that were comprehensible and highly relevant. These pages were translated into English using Google Translate and analyzed with the 162 English pages above.

Next, we conducted a second round of analysis to identify off-topic text that could be excluded from our intensive coding process. We examined all Meta-Wiki/Wikipedia namespace pages that describe the functionalities or moderation guidelines of FlaggedRevs or Pending Changes and  %
 the Discussion/Talk/Village pump pages. These pages were, on average, 25,000 tokens long, with hundreds or even thousands of user comments expressing their experience with the system. Please note that, because wiki pages included both words and markup, word count is not an effective measure of length. A better measure is tokens, where tokens are words or fragments of markup. 

Because we filtered out all  Phabricator pages that describe minor code maintenance tasks, our analytic dataset retained only tasks where the original poster asked for permission to deploy FlaggedRevs on their wiki or asked for a major reconfiguration of the software. For example, we kept the Phabricator Task where an administrator from Indonesian Wikipedia asked WMF's software developers to ``reconfigure the current FlaggedRevs setting from full application to the so-called `English approach' [i.e. Pending Changes]''.\footnote{\url{https://phabricator.wikimedia.org/T268317}} Finally, because Special namespace pages are generated automatically, we only gathered relevant statistical reports and excluded most pages from our coding process. In total, the core dataset that was used in subsequent analyses included 67 documents with a total of 1,312,878 tokens.

\subsection{Semi-structured Interview Study}

To complement our archival data, we sought the perspective of people who used and maintained the software. With the approval of the Institutional Review Board at both participating institutions, we conducted seven semi-structured interviews with individuals who used FlaggedRevs and/or who demonstrate intimate knowledge about its operation and maintenance. We also identified potential candidates by using our QFA dataset to identify people who recently conducted moderation work using FlaggedRevs or had been involved with FlaggedRevs-related tickets on Phabricator. Subsequently, we used a snowball method to ask our interviewees to introduce us to others with direct knowledge about FlaggedRevs who might be willing to be interviewed by us in English. We conducted new interviews as we analyzed data until the interview data helped us fully triangulate our findings.

Table \ref{tab:intervew_study} describes our interviewees' gender, role in relation to FlaggedRevs, the Wikipedia project to which they primarily contributed, and the interview length. Five of our interviewees are administrators in four different wikis that deployed FlaggedRevs: Indonesian Wikipedia (I1), Farsi Wikipedia (I2), English Wikipedia (I3, I4), and Hungarian Wikipedia (I7). Furthermore, some interviewees also worked for WMF in teams including the Movement Strategy, Moderation Tools, and Legal teams. To avoid deanonymizing our research subjects, we do not attribute staff to particular positions in Table \ref{tab:intervew_study}. Moreover, we interviewed a WMF software developer working for the Anti-Harrassment team (I5). Although this person has not personally used FlaggedRevs, they demonstrated extensive knowledge about WMF's approach to content moderation. Finally, we interviewed a user without administrator status who contributes and moderates content using FlaggedRevs in Polish Wikipedia (I6). All interviews were conducted in English via video call on Zoom. We obtained verbal consent from our participants to record these interviews and to transcribe and analyze them later. Each participant was offered \$25 compensation. After our coding process, all records and documents that contain participants' personal information were anonymized and/or deleted.

\begin{table}
\caption{\textbf{Interview participants' gender, role, community, and interview time in minutes.}  ``—'' indicates that the participant does not primarily contribute to a particular Wikipedia project.}
\centering
 \begin{tabular}{l l l l l} 
\hline
 \textbf{ID} & \textbf{Gender} & \textbf{Role} & \textbf{Community} & \textbf{Interview Time} \\ [0.5ex] 
 \hline
 I1 & Male & Administrator/WMF employee & Indonesian Wikipedia & 62m\\
 I2 & Male & Administrator/WMF employee & Farsi Wikipedia & 50m\\
 I3 & Male & Administrator/WMF employee & English Wikipedia & 45m\\
 I4 & Non-binary & Administrator/WMF employee & English Wikipedia & 37m\\
 I5 & Female & WMF employee & - & 32m\\
 I6 & Male & Moderator & Polish Wikipedia & 31m\\
 I7 & Male & Administrator/WMF employee & Hungarian Wikipedia & 48m\\
 \hline
 \end{tabular}
 \label{tab:intervew_study}
\end{table}

In interviews, we first established the background and experience of each participant, and their experience with FlaggedRevs, as well as to other moderation tools. Further, we asked participants detailed questions about the timeline of FlaggedRevs deployment in their communities as well as their experience with, and assessment of, the tool. We asked participants to describe their personal experiences with and opinions about FlaggedRevs and the reactions of their communities overall. One participant used only FlaggedRevs for their moderation work (not any other tools), and one participant never used it personally. Although both acknowledged that they could not make informed comparisons between FlaggedRevs and other moderation tools, they answered other questions regarding their content moderation work on Wikipedia.

We also sought to understand the decision-making process behind the deployment and maintenance of FlaggedRevs in each community. As some of our interviewees worked for the WMF, we asked them about WMF's moderation policies and procedures used by WMF to coordinate with and provide technical support to volunteer administrators. Sometimes, interviewees would provide us with documents that were relevant to our questions. We added these to the corpus of textual data described in the previous section. Our interview transcripts were combined with the full corpus of documents for qualitative data analysis.

\subsection{Qualitative Data Analysis}

We conducted our qualitative data analysis using the constant comparative method as described by \citet{charmaz2014constructing} in her book titled ``Constructing Grounded Theory''. The constant comparative method helps researchers derive a set of patterns and theories from a systematic analysis of data and is common in qualitative studies in social computing~\cite{10.1145/3274313, 10.1145/3359328, 10.1145/3568491}. %

First, we performed inductive open coding; the first author did line-by-line coding of both the documents collected in the QFA process and the interview transcripts. Subsequently, we began to form themes by identifying the most significant and/or frequent codes gathered earlier. These themes, in turn, were connected to form a set of broader themes.

For example, notable themes such as ``workload imbalance'', ``disapproval of higher barrier to entry'', ``struggle to reach consensus'', and ``creation of social hierarchy'' helped us tweak our interview script to confirm and dig deeper into these issues. We repeated this process until we could no longer discover new themes from the coding process. Finally, we constructed detailed memos about our derived themes, accompanied by textual evidence from both our interviewees and Wikipedia members expressing their opinions online. The findings reported in this paper reflect the most important themes we identified. 

\section{Findings}
\label{sec:findings}
Wikipedia contributors typically understood that the potential cost of operating FlaggedRevs would be high. In 2009, when English Wikipedia discussed the idea of adopting the system, the community concluded that they could not realistically moderate edits made to the hundreds of thousands of articles in English Wikipedia with the editor resources available at the time. In the words of one editor, ``aggressive use of [FlaggedRevs' features] was rejected by the English Wikipedia community, essentially for too drastically reducing the ability of anyone to edit Wikipedia'' \cite{fr_patrolled_revisions}. Instead, the English Wikipedia community decided to organize several pilot runs with a substantially limited configuration of FlaggedRevs. Further, the system was scaled back by limiting the number of articles  affected by the system. In practice, only heavily vandalized articles are subject to FlaggedRevs on English Wikipedia, while most are not.

\begin{figure}
 \centering

  \includegraphics[width=\textwidth]{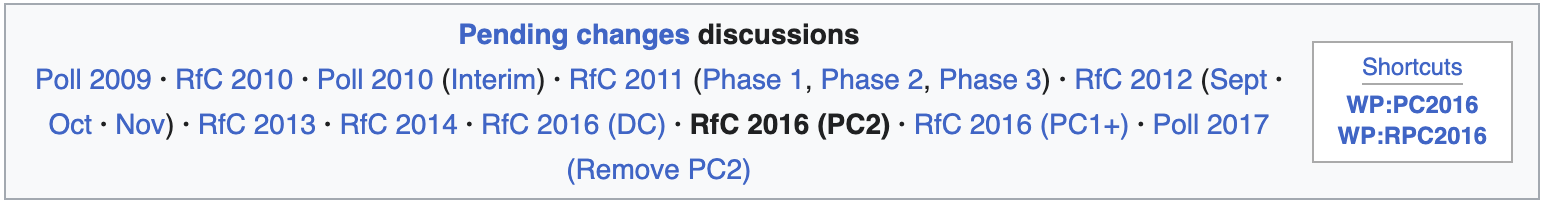}
  \caption{List of discussions and polls regarding the implementation of Pending Changes on English Wikipedia.}
 \label{fig:pc_discussions}
\end{figure}

The English Wikipedia community spent the next eight years trying to improve and integrate FlaggedRevs into their workflow and went so far as hiring ``at least one contractor [...] to tweak and test the configuration, as well as development work involving no less than 8 WMF persons'' \cite{flaggedrevs_main}. Figure \ref{fig:pc_discussions} shows a list of discussions and polls on how to implement the scaled-back version of FlaggedRevs for English Wikipedia from 2009 to 2017. English Wikipedia's scaled-down version was ultimately named ``Pending Changes'' and it is typically referred to that way by the community. Although Pending Changes uses the exact same code base as FlaggedRevs, it is often thought of and described by Wikipedia community members as a different system altogether.

Next, most active community members in communities that deployed FlaggedRevs were mentally prepared to trade off the instant gratification and ease of access that Wikipedia contributors are used to with a new moderation policy emphasizing quality because they believed it would lead to higher degrees of trust from readers. As time went on, however, some community members started to realize many unforeseen challenges occurred with the adoption of the new system that could not be addressed easily. Our analysis revealed \textit{eight} primary challenges that we have grouped into three areas, namely, \textit{community challenges},\textit{ platform and policy challenges}, and \textit{technical challenges}.

\subsection{Community challenges}

\subsubsection{Conflicts with existing social norms and values.}

FlaggedRevs subjects contributions by not-yet-trusted users to mandatory review, only displaying the latest \textit{approved} version of the article to general audiences.\footnote{\url{https://meta.wikimedia.org/wiki/Flagged_Revisions}} While language editions were eventually allowed to choose whether they wanted to display the latest version of the article by default (even if it was not yet approved), readers were always notified as to whether the latest version of the article had been reviewed. To deploy FlaggedRevs, two key questions must be answered---\textit{who} can approve edits and \textit{whose edits} would need to be approved. 

Wikipedia historically prided itself on being a platform that welcomes everyone to contribute, with or without an account. Outside of a small number of users with administrative rights, unregistered and registered users can perform the same editing and moderating functions on most articles, including the ability to revert others' contributions. The deployment of FlaggedRevs---a prepublication moderation system---fundamentally changed this by creating a new social hierarchy within the community.
FlaggedRevs creates two \textit{explicit} user classes: a user class who receives  \textit{autoreview}  (i.e., the right to have their contribution automatically approved and published), and a \textit{reviewer} class who receive both \textit{autoreview} and the right to review and approve untrusted edits. 

In the survey about the deployment of FlaggedRevs on German Wikipedia, a user pointed out that ``the process of [Flagged Revisions] promotes the hierarchization of Wikipedia and is a step towards a closed society'' \cite{german_wiki_poll}. Another user concurred that ``this class society or one-sided surveillance on a hierarchical basis did not previously exist in Wikipedia, and all in all it has worked up to now!'' Overall, at least 10 mentions of some form of the word ``hierarchy'' are found in the survey results, making it one of the main reasons users argued against the deployment of FlaggedRevs. This same sentiment was repeatedly expressed after each trial run of Pending Changes on English Wikipedia between 2011 and 2012. For example,

\begin{quote}
    My main concern with Pending Changes is that it essentially requires a new class of users (Reviewers). I really do not think this fits well with Wikipedia as this introduces more user hierarchy and may act to discourage users from becoming active in the community by putting up barriers on the kinds of additions/help users can make (English Wikipedia user).
\end{quote} 

These aforementioned quotes highlight the difference between two groups with opposing viewpoints. One user group thinks that giving some users reviewing rights will create a user hierarchy and increase barriers to entry. The other thinks that doing so is necessary to ensure high-quality articles. In Orlikowski and Gash's terms, the conflicting perceptions of the tool and its subsequent impact on the community' structures indicates the existence of an incongruent technological frame.

Others listed additional reasons that the introduction of a user hierarchy would be problematic by expressing concerns about the potential for abuse from the newly created reviewer class who might use their power to gatekeep certain article pages by rejecting any contribution made by the less trusted group, regardless of its quality. This was noted in the summary of a consensus vote on whether Pending Changes' problems outweighed its benefits \cite{pc_RfC_2012}:

\begin{quote}
    Some Position 1 [Opinions that negative aspects of Pending Changes outweigh the positive] supporters felt that the implementation of Pending Changes would put in place a hierarchy of editors where one hadn't existed before. Some of these arguments were based on the assumption that those given ``reviewer'' status would use it to force through their preferred content or that those able to turn on the ``reviewer'' right for other users would hold it hostage. (English Wikipedia user)
\end{quote}

Although not all platforms with decentralized governance will react so strongly to a new social hierarchy being created within their community, Wikipedia users felt that doing so went directly against the existing norm of allowing anyone to edit freely. While  everyone is still allowed to edit under the new system, an English Wikipedia editor characterized the system as ``crush[ing] the wiki principle of `free participation' and relegates new and [unregistered users] to second place.'' 

Another way this broader concern was voiced was in terms of how FlaggedRevs changed the reward system at play in Wikipedia and created a higher barrier to entry. According to interviewee I1, applying FlaggedRevs to all articles creates ``too much of a high barrier to entry for editing,'' and is one of the reasons members of Indonesian Wikipedia voted to use the scaled back configuration they called Pending Changes. Furthermore, I2 pointed out that FlaggedRevs ``takes away the instant gratification, and people don't feel super excited about staying there and being in Wikipedia and contributing anymore.'' Indeed, instant gratification can be seen as a form of reward for volunteers, who do not receive any tangible compensation for their work. For new users who have yet to experience a sense of belonging in the community, waiting days, weeks, or months for their contribution to be accepted might give them little reason to stay.

Peer production platforms often rely on low participation costs to promote growth \cite{hill2021hidden}. Being concerned that a new technology that changes the reward systems for new users could negatively affect growth is reasonable. Wikipedia was experiencing a steady decline in active users after years of growth around the same period in which FlaggedRevs was initially deployed \cite{halfaker2013rise}. I3 explained that this contemporaneous decline became a top priority that Wikipedia was trying to address: ``It was around the time that editor numbers plateaued and we first saw the graphs with edit numbers declining, and there was kind of panic around `OK, we need to think about retention or Wikipedia is going to die.'''
Sharing the same concern, an administrator from Hungarian Wikipedia (I7) explained that he and other Hungarian Wikipedia administrators turned off FlaggedRevs for a period of time in 2017 to test whether the user retention metrics would improve. Ultimately, they reported seeing no noticeable effect \cite{huwiki_report}.
Recent scholarship has suggested that FlaggedRevs' deployment is unlikely to have made a major contribution to these dwindling growth rates \cite{chau_tran_flaggedrevs}.

\subsubsection{Unclear moderation instructions.}

For any moderation system to be successful, moderators must understand the purpose of the system. For FlaggedRevs, this was largely done by creating and enacting written guidelines. As FlaggedRevs is designed to be flexible and give communities more power to decide how to moderate their content, no single comprehensive guideline exists. For example, a user from the German Wikipedia pointed out that little to no information exists about how to judge and approve an edit:

\begin{quote}
There seems to be a great deal of ambiguity as to what exactly the requirements are for marking an article as viewed. The only wording that can be found, that the article should not contain ``obvious willful defacement (vandalism)'', is too vague and, in my opinion, does not fulfill the actual meaning of the review. On the one hand, the question arises as to what ``obvious'' vandalism is, especially in the case of the ``deliberate falsification of information'' variety mentioned under Wikipedia:Vandalism.\footnote{\url{https://de.wikipedia.org/wiki/Wikipedia:Vandalismus}} (German Wikipedia user)
\end{quote}

\noindent In the mind of this user, instructions might guide reviewers through difficult moderation decisions, limiting their reliance on personal viewpoints and enabling more equitable and uniform reviewing practices. Guidelines could enable congruent moderation frames across reviewers and between editors and reviewers.

However, creating such guidelines reflected a challenge. Insight from I6 from Polish Wikipedia reveals that the community struggled to agree on rules and criteria to define vandalism:

\begin{quote}
    I think the first problem is criteria. So there was always a discussion among Wikipedians, and it is an ongoing discussion about what we should and should not flag. So the tool asks if there is any obvious vandalism, but others are using and abusing this tool to require edits to have more stringent requirements in terms of quality. So I guess it should come with clear and well-discussed consensus of how we go about moderating content. What criteria should we apply? (I6)
\end{quote}

In a similar vein, a user from English Wikipedia raised the issue of what they called a ``credibility gap.'' This issue describes how even if a reviewer is making a good-faith effort to review an article, and their attempt might be impacted by their limited knowledge of the topic:

\begin{quote}
    This idea only really works with obvious vandalism, and obvious vandalism is---well---as obvious to the general public as it is to the `approved' editor/reviewer. The genuine credibility gap in Wikipedia is when an expert (or at least someone knowledgeable about a subject) is reading a topic they know something about and they see that there are mistakes. Most approved editors/reviewers will not have that level of understanding of a topic or they'd already be editing the article, so most articles will either be unflagged or, more likely, flagged incorrectly, which is even more damaging to our credibility. (English Wikipedia user)
\end{quote}

\noindent Throughout our dataset, Wikipedia users appeared to disagree on whether FlaggedRevs was for checking an edit for accuracy, quality writing, sloppy mistakes, or obvious vandalism. 

One user argued against using FlaggedRevs for superficial spot checks arguing that  incorrect information might be added to an article that might go unnoticed for months before being corrected. If an edit is accepted simply because it contains no obvious grammatical mistakes, it will be published and will negatively affect the quality of the article. Moreover, for some German Wikipedia users, a moderation system that holds up edits, but only for superficial checking, would be (and was) ineffective and that ``there should be a kind of quality stamp that the article is 100\% correct in its current form.''
Further, counter-arguments tended to emphasize that vandalism attempts are obvious and that FlaggedRevs could help with ``spoiling the fun for vandals since they can no longer immediately admire their mindless edits.''

Moderators might not demonstrate the time or knowledge to check every edit for quality and factualness. An English Wikipedia user expressed concern about the responsibility of reviewers:

\begin{quote}
    What exactly is the PC reviewer's responsibility to prevent such additions? If a vandal writes that the daughter is a porn star, followed by a reference to an offline book or magazine article, is the PC reviewer supposed to look up and find out if it's a real book? (English Wikipedia user)
\end{quote}

\noindent Furthermore, needing to spend time and effort reviewing all edits carefully could overwhelm reviewers. The arguments went back and forth on German Wikipedia without any clear consensus for a moderation guideline \cite{german_wiki_poll}. After administrators gathered feedback for the system, one of the concluding points stated that even though people disagreed on the substance of the guidelines, they agreed that they wanted ``to see a concrete policy laid out before it is used, as they believe it could potentially be misused if such a policy doesn't exist.'' This is consistent with Orlikowski's comment that the lack of explicit procedures and policies around a new system highlights the difficulty of enforcing policies \cite{orlikowski92fromnotes}. 

\subsection{Platform and policy challenges}

Although the WMF demonstrates some overarching policies, it largely allows communities to set rules within their sites. The WMF exhibits anti-harassment and moderation tool teams formed in 2017 and 2020, respectively. In interviews, we learned that neither group is responsible for developing any general content moderation system or process. When FlaggedRevs was first developed, neither team existed  

\subsubsection{Lack of technical support from the governing body.}

Through interviews, we learned that after FlaggedRevs was created by a group of third-party contractors and developers at Wikimedia Deutschland, it was up to each language edition to decide whether to use the system and how to configure it. After contractors moved on from their work, WMF asked volunteer developers to maintain the tool.  This hands-off approach resulted in a lack of technical support and code ownership. Participant I2---who is currently the only person in charge of maintaining Flagged Revisions' code base---explained that the lack of ownership is not unique to this project:

\begin{quote} 
    Historically, the FlaggedRevs is developed also within the Wikipedia infrastructure, but, in turn, no one really wants to maintain it anymore, so it's really huge code that is being left alone to be patched by the volunteers (I2).
\end{quote}

  The lack of technical support is detrimental to the longevity of any  software, regardless of how well it performs. As FlaggedRevs is deployed at the wiki level, tweaking the software is not easy for language editions. Therefore, they must get a rough consensus within their community---a requirement for technical administrators at WMF to accept and act upon any request. Due to this multi-step decision process, problems with FlaggedRevs could often not be addressed in a timely manner. I1  explained that this abandonment of projects was not an isolated case:

\begin{quote}
    FlaggedRevs is one of them [...] There were one or two others, some of these content moderation tools, and they worked on them for a while and the communities have appreciated them. And then, they got dropped for one reason or another, and the sort of priority shifted to other work and they basically sat there unmaintained for the better part of 10 years and so at this point, the effort in sort of getting a product team to learn about this very old code base to renew it, maybe to have to rewrite it from scratch, is a pretty tall ask [...] (I1).
\end{quote}

This quote raises the question of what potential avenues exist for Wikipedia community members to request technical support. I6, who is a project manager at WMF, shared with us that the organization tends to take a hands-off approach, even for the tools that they developed:

\begin{quote}
     The Wikimedia Foundation does not play a big part in those conversations [regarding who is in charge of maintaining the software]. My understanding is that the community itself organizes around maintaining these goals, when the volunteer chooses to step down. For maintenance of the tool, they would usually ask somebody else to step up and take that role, and there are at times they are just groups of people who are maintaining a slate of tools because they have been around, and they know the tech a little bit, so there's always self-organization there. (I5)
 \end{quote}
 
Most of our interviewees confirmed that community members would typically request technical support for FlaggedRevs through the Community Wishlist survey \cite{community_wishlist}. In this process, users can submit a request for support during the proposal phase and, depending on the number of votes from other community members the request receives, the Community Tech team within WMF will work with the community to find a feasible solution before starting work.

There are limits on the kind of requests that users can make through this process. Additionally, the Community Tech team will often decline a request if they feel they cannot complete it.  As the team is quite small, they can only accept a handful of small-scale requests per year. Software that needs a substantial code base update (i.e., more than a year of work), like FlaggedRevs, was described as being very unlikely to be worked on by the team at all \cite{community_wishlist}. In this sense, responsibility for maintenance of FlaggedRevs is simply not clear. As a result, language editions using FlaggedRevs must not only ensure that they have enough reviewers to carry out reviews, but must also be sure that they can access technological support. Frequently, they could not rely on either.

\subsubsection{Bureaucratic hurdles.}

The process of deploying FlaggedRevs on English Wikipedia is an example of the bureaucratic hurdles that online communities face when they attempt to restructure their moderation practices. As we mentioned in  §\ref{sec:empirical}, the English Wikipedia community quickly realized that they could not deploy Flagged Revisions at the scale of the entire encyclopedia. So, between 2009 and 2011, the community held multiple discussions and polls to identify ways to change the default configuration of FlaggedRevs in ways that would work best for English Wikipedia. This included discussion around the threshold for articles to be protected by the tool, assigning user roles, and so on. These configuration changes were so substantial that the English Wikipedia community referred to the system with an entirely new name (Pending Changes) to avoid confusion with the original configuration used on German Wikipedia.
After these discussions, community members held numerous trial runs between 2011 and 2016 to test the effectiveness of the reconfigured system and to gather feedback. While they eventually reached a consensus on how to use the system, the prolonged process led to frustration:

\begin{quote}
    This project is displaying some weird kind of \textit{continuismo} that just can't be justified by saying ``They might fix it any time, who knows.'' The problem is, when the community can't stop an experiment despite clear resolution to do so, it creates general resistance to any new experiments. (English Wikipedia user)
\end{quote}

This situation was hardly unique to English Wikipedia. In 2017, Indonesian Wikipedia members decided to change Flagged Revisions' configuration on their wiki to be more like Pending Changes \cite{idwiki_fr}. However, even after consensus was reached within the language community, they needed to have WMF staff execute the configuration change. I1 shared that the request was not approved until 2021 due to some communication and translation issues: 

\begin{quote}
    The vote was back in 2017, but the developer community---which has the ability to execute these changes---was a bit confused. I think it's mostly a language gap problem of what the vote decided to do. And only four years later, when a bunch of people are saying ``hey, you haven't responded to a request for four years'. And they said `What do you really want?''. ``We want Pending Changes'. ``Oh, you should have said that four years ago''. 
    (I1)
\end{quote}

\noindent This quote reveals the multiple steps required to execute a decision about moderation change in this environment. It also indicates the importance of aligning technological frames between stakeholders, in this case, users and WMF developers. %

\subsubsection{Lack of cross-community empirical evidence on the effectiveness of the system.}

Additionally, the decentralized nature of  Wikipedia resulted in the absence of cross-community performance measurements that could guide the decision-making process about the use of FlaggedRevs. Despite having numerous Request for Comments (RfC) pages collecting feedback, no in-depth analysis occurred to provide an assessment of how well the new moderation system performed. Many users provided their own opinions regarding the tool but lamented the lack of empirical evidence. On an RfC page, one English Wikipedia user complained that ``after 2 months of a real trial and 5 months of \textit{fait accompli}, we still have no real data on the influence of [Pending Changes] on edits, editor conversion, and vandalism.'' Even from the perspective of the platform governing body, no system was in place to gather performance metrics for the moderation system across wikis. 

Participant I3, who is on the Wikipedia Moderation Tools team, shared his personal struggle in understanding the effectiveness of FlaggedRevs: 

\begin{quote}
    I really struggled to find numbers on this so it piqued my curiosity when we were looking at this and I wondered, first of all, was there any data? What were these decisions being based on? And second of all, internally, how are we thinking about this? Because the Foundation has explicitly said ``We don't want to deploy FlaggedRevs anymore because we think it's having a negative impact,'' but when I asked people like who either involved in the decision all around the decision, there was just kind of this feeling that it was probably bad and no one could really point to like ``Here was the number we saw that was bad''. Everyone just said ``well, you know, obviously it's bad'' or, like ``you know, there's a feeling it's bad'' (I3).
\end{quote} 

Without an evaluation of the tool, it is easy to think that a system liked FlaggedRevs might negatively affect contributor growth. 
This lack of cross-community performance metrics is also a problem with AutoModerator on Reddit, where this type of data has been described as being potentially ``vital for human moderators to understand how different configurations of automated regulation systems affect the curation of their sites'' \cite{jhaver_automoderator}.

\subsection{Technical challenges}

Our analysis suggests a range of technical challenges may harm user experience, which, in turn, may hurt the system's long-term prognosis.

\subsubsection{Scalability and production misalignment.}

Scalability is one of the primary concerns moderators voiced with nonautomated prepublication moderation tools like FlaggedRevs. Because they require moderators to approve pending contributions, unreviewed contributions can easily snowball if moderators cannot efficiently handle them. This problem is partially addressed by the ability to reward active registered users with a promotion such that their work will be automatically approved. However, as time goes on, backlogs can become effectively unrecoverable for many wikis with FlaggedRevs enabled. We frequently read and heard these complaints.

I4 shared their struggle in dealing with the backlog problem, saying:

\begin{quote}
    [...] so any random page with an IP makes an edit, someone must review it, and we don't have the capacity of volunteers who can go through all these backlogs and review everything and judge if it's good or not (I4).
\end{quote} 

We collected data highlighting the severity of backlog issues from each wiki's ``Statistics on Flagged Revisions'' page, which we present in Table \ref{tab:waittime} \citep[e.g.,][]{german_fr_waiting, polish_fr_waiting}. While not all wikis maintain this statistics page, we collected the mean and median wait times (as well as the 35$\mathrm{th}$, 75$\mathrm{th}$, and 95$\mathrm{th}$ quantiles) for ``pages with edits currently pending review'' under the FlaggedRevs system in 11 language editions. We observe that the distributions are skewed, with a large difference between mean and median. The 95$\mathrm{th}$ percentile is very large on many wikis which suggests that there are articles with unreviewed contributions that no one has looked into for dozens to hundreds of days. 

\begin{table}
\caption{\textbf{Reported wait time for ``pages with edits currently pending review'' on 11 wiki language editions.} We assembled the table by pulling data from each wiki's statistics page when provided \citep[e.g.,][]{german_fr_waiting, polish_fr_waiting}. Due to the huge time difference ranging from minutes to months, we convert the time unit into \textbf{hours}.
}

\centering
 \begin{tabular}{l l l l l l} 
\hline
 \textbf{Wiki} & \textbf{Average} & \textbf{35th Percentile} & \textbf{Median} & \textbf{75th} & \textbf{95th} \\ [0.5ex] 
 \hline
 Albanian&826&47&48&794&4159\\
 Alemannic&1&\textless1&\textless1&2&7\\
 Bosnian&158&\textless1&\textless1&6&1530\\
 Belarusian&6&\textless1&2&12&26\\
 Esperanto &358&108&42&528&874\\
 Georgian&34&6&21&54&145\\
 German &10&\textless1&2&8&24\\
 Hungarian&26&\textless1&4&22&127\\
 Polish&10&\textless1&2&13&43\\
 Russian&478&69&326&684&732\\
 Ukrainian&1493&25&648&3415&4073\\
 \hline
 \end{tabular}
 \label{tab:waittime}
\end{table}

One potential explanation for this skewed distribution relies on understanding user moderation behavior on Wikipedia.
Evoking ideas of technological frames, users on one Pending Changes RfC page shared their thoughts on the existing moderation practices of users of Wikipedia and the implication of FlaggedRevs:

\begin{quote}
    My personal experience has shown that active Wikipedians fall into two basic groups: one group takes care of the pages they watch and goes so far as to check every single change to the articles, the other group watches hundreds, if not thousands, of pages and checks superficially and not in terms of content, if only for capacity reasons. For the latter, FlaggedRevs is optimal for their intention, they do not have to and do not want to carry out a content check if it does not appear to them to be vandalism. The first group proceeds differently here, since every change is checked anyway and always in terms of content---regardless of whether Flagged Revisions is in Wikipedia or not (English Wikipedia user).
\end{quote}

\noindent This user suggests that there are some moderators who regularly patrol each community's ``Recent Changes'' page to check recently submitted edits and to approve or flag edits in the process.
Other editors simply want to focus on editing articles that correspond to their interests.
If the first group of users misses an edit during a high-volume cursory review (e.g., using a feed of Recent Changes), it may take a very long time for a user who conducts in-depth reviews to see the change---especially for less popular topics without devoted reviewers.

\subsubsection{Design issues.}

By default, FlaggedRevs ensures only the latest version of an article that a moderator has reviewed will ever be shown to most readers. Users sometimes refer to this as the ``stable'' version. If an editor later makes a contribution without \textit{autoreview} rights, it will be subjected to review and can only be seen by users who have logged in. All contributions that come afterward will also be marked as unreviewed. This will result in a ``fork'': two versions of the article that exist in parallel. The presence of two versions makes collaboration between users more difficult. Participant I2 described this forking issue:

\begin{quote}
The way that FlaggedRevs works is that, if you imagine two revisions just being edited on top of each other. So once an edit that comes in from an IP [unregistered user], it must be reviewed. So it starts to fork. The first revision will be called the stable version [...]  even after that IP edited something, if anyone edits on top of that, it's still not going to be shown because the IP edit has not been reviewed yet (I2).
\end{quote}

\noindent As the backlog of unreviewed works grows, articles display an increasingly outdated stable version. While the stable version of an article might be high quality along some dimensions, if the information  is heavily outdated, users might not be interested in reading it. This is an issue for time-sensitive topics, like breaking news, which make up a significant portion of all traffic to Wikipedia \cite{keegan2012breaking}.

Some users suggested that FlaggedRevs should show everyone the latest version of an article, even if it has not yet been reviewed, as long as readers are notified that the revision has not been vetted yet. While German Wikipedia eventually voted to keep displaying the stable version by default \cite{german_wiki_poll}, some smaller communities asked developers of FlaggedRevs to allow them to reconfigure the tool to display the unstable version. They felt this was better because they did not have the resources to review articles in a timely manner. This was approved and added as a setting in FlaggedRevs. However, displaying the unstable version by default means vandalism is visible in a way that effectively eliminates almost any of the purported benefits of the FlaggedRevs system and further reduces the incentive for established users to engage in the work of reviewing changes.

\subsubsection{Compatibility with other moderation tools.}

Volunteer moderators often create tools to support their work. For example, many experienced Wikipedia users use \textit{Huggle} and \textit{Twinkle}, tools that attempt to detect  and revert vandalism. When an article receives an uptick of vandalism attempts, administrators can enable \textit{protection} or \textit{semiprotection} mode to temporarily lock an article and restrict contributions from some, or even all, users without administrator rights  \cite{protection_policy}.
There are hundreds of tools and bots that Wikipedia users have created to help with moderation work.
As a prepublication moderation tool, FlaggedRevs changes the workflow of moderators. For example, it overlaps and conflicts with the widely used practice of \textit{semiprotection} \cite{protection_policy}. In the default settings that affect all articles within a language edition, FlaggedRevs  renders semi-protection redundant.

\begin{figure}
\includegraphics[width=\textwidth]{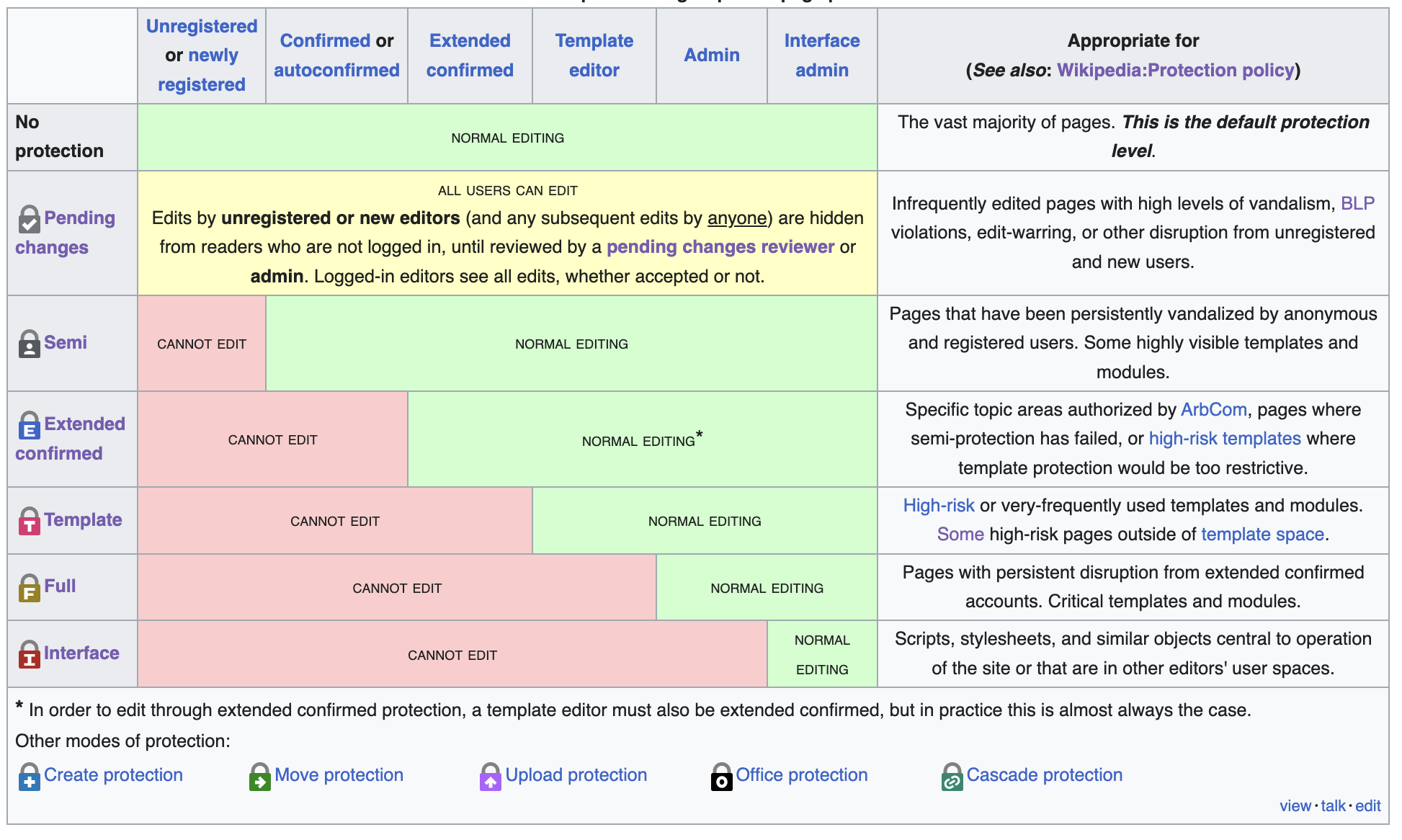}
\caption{Page protection levels and corresponding user accessibility on English Wikipedia.}
\label{fig:protection_levels}
\end{figure} 

Recognizing this redundancy with semi-protection, the English Wikipedia community attempted to find a reasonable use case for FlaggedRevs. The community eventually agreed to use the system for ``infrequently edited [articles] with a high level of vandalism'' before considering the use of other more restrictive protection modes. Figure \ref{fig:protection_levels} further illustrates different protection levels and the corresponding protocol to be triggered when an article page is deemed at risk of vandalism. For example, when an infrequently edited page experiences a higher-than-usual level of vandalism, administrators can turn on Pending Changes protection mode to ensure edits by unregistered or new editors are hidden by default. For pages that are more frequently attacked by vandals, even after enabling Pending Changes, administrators can consider higher protection modes like semi-protection or a protection level called \textit{extended confirmed} to fully restrict editing for a broader class of editors, not only those who are unregistered or new.

Although Pending Changes can be deployed in a way that complements moderation tools like Twinkle and Huggle, this did not satisfy all Wikipedians' concerns. Some argued that Pending Changes only helped pages after they had been vandalized frequently, while the goal and benefit of the system was to prevent the negative impact of any vandalism attempt. This concern was similarly echoed by the interviewee from Hungarian Wikipedia (I7):

\begin{quote}
    The way [English Wikipedia] uses [Pending Changes] is basically as a kind of page protection. You can protect a page when it becomes problematic like it's in the news, or there is an influx of a lot of editors who don't understand the rules and make problematic edits. Instead of normal protection, which is just completely excluding users or new users or anonymous users from the page, you can do this pending protection thing, which means that they can still edit, but the Flagged Revisions dynamics get applied to their edits. I think this is sort of concern that only large wikis have when you get lots of edit from new users who come to a page because it's controversy or current topic in some way. This is not happening too much in smaller wikis (I7).
\end{quote}

\noindent This quote highlights the differences between the effect of the system for smaller and larger wikis. It indicates that besides challenges from tool incompatibility, incongruent technological frames might also be a consequence of wiki size.
Regardless of the argument, applying any system that overlaps with existing systems in terms of functionality remains a challenge. It is difficult to convince communities to embrace a disruptive technological change like FlaggedRevs without evidence that it can outperform existing systems. FlaggedRevs often fell short in this way.

\section{Limitations}
\label{sec:limitations}

Our data collection and findings are heavily limited by language barriers. FlaggedRevs are deployed in 25 language editions. Although our team speaks several languages, we cannot read or understand many of the language editions on which the system is deployed. Even with the help of Google Translate, we could not adequately translate some of the text we collected. We attempted to address this issue by reaching out to several non-English wikis who use FlaggedRevs for our interview study. However, we were unable to recruit many, perhaps due to the same language barriers. The challenges faced the communities mentioned in our findings might be extremely different from the challenges that other communities experienced, especially smaller communities built around relatively under-resourced languages.

Another limitation of our study is the time difference between our textual data and our interviews. Although our methods allow us to observe and gather context-sensitive evidence, analyzing users' attitudes toward FlaggedRevs at the time they were discussing how to best use it is different from asking our interviewees about their own analysis of the system long after its adoption was halted. For example, we know that at the time it was being deployed in new wikis, many users were concerned that FlaggedRevs might be contributing to the steady decline in user base. Moreover, recent analyses by \citet{chau_tran_flaggedrevs} suggest that the system did not contribute to lower user retention. Given these results and their own experiences, our interviewees' view of FlaggedRevs tended to be much more neutral than many of the voices in our archival data with little sense of urgency. In their mind, FlaggedRevs is merely one obsolete moderation tool among many. If we interviewed the same people at the time of its peak adoption, our interviewees might have shared different opinions. Furthermore, we believe that some of the findings from our interview study might suffer from hindsight bias. Although a potential limitation, we see these differences as important sources of insight in and of themselves and opportunities for triangulation. 

Some of the challenges highlighted in this paper might be unique due to the structural properties of Wikipedia, the relationship between WMF and the various language communities, and the characteristics of FlaggedRevs. Wikipedia, with the desire to be a more credible encyclopedia, found itself wanting to drastically restructure moderation practices without fully understanding the consequences. Many decentralized platforms relying on community-based moderation would not expend such considerable effort in developing and experimenting with new technologies.

\section{Discussion}
\label{sec:discussion}
Our findings suggest that FlaggedRevs is more complicated and more interesting than a simple failed technology. If FlaggedRevs was more trouble than it was worth, many language editions would have decided to remove it completely. Instead, wikis continue to employ the system. Some expend considerable effort to tweak it to fit their needs; however, many use it as-is. German Wikipedia, the second largest in number of articles, continues to run the system at full scale effectively.
The reasons for wikis persisting to use the tool is unclear; however, we know that communities that use it do so at some cost in terms of time and labor, and they seem willing to pay that cost.

Three decades later and in very different technological and organizational contexts, our findings are roughly in line with Orlikowski and Gash's observations. Similar to Orlikowski and Gash, we find that features of FlaggedRevs that were incompatible with Wikipedia’s culture, policies, and reward systems prevented it from being fully deployed. 
Despite the broad similarities, certain important differences and extensions in our findings also exist.

Next, we found that one of Orlikowski and Gash's dimensions of technological frames, ``philosophies towards technology,'' was even more salient in the more fluid communities that comprise Wikipedia than in Orlikowski and Gash's formal organizational setting. While Orlikowski and Gash discuss personal and organizational philosophies, we find that subcommunities' beliefs and assumptions (e.g., around new contributions and hierarchy) led to incongruent frames in regard to FlaggedRevs.
Orlikowski and Gash show how introducing technological change to a community is challenging. Our work also suggests that doing so across different communities with different sizes, cultures, structures, and needs--as well as those that interact with each other and discuss the technology in question---is much more complicated. 

Another difference stems from the fact that our findings indicate that FlaggedRevs was thoroughly considered but never adopted in multiple communities. This suggests that incongruent frames appeared in what Orlikowski and Gash call the ``pre-intervention'' phase before the technology was adopted. In fact, we found that some of the questions about how FlaggedRevs reflects or affects community norms and values led to long-term conflicts \cite{halfaker11dontbite}, indicating that building congruent technological frames may be important at earlier stages in contexts practicing more participatory forms of governance.

Further challenges stem from the polycentric nature of Wikipedia's governance. The tools available to Orlikowski and Gash's managers for aligning frames---as limited as they were---make that context look simple compared to the FlaggedRevs setting. Moreover, the successful implementation of FlaggedRevs requires alignment between stakeholders including WMF, language editions, committees, developers, and operators \cite{jhaver2023decentralizing}. Our findings indicate that achieving shared technological frames among all these stakeholders was often impossible. 

Our results suggest that frames became aligned in ways that caused the system to be effective by empowering subcommunities to figure out how they wanted to use the tool and to conduct their own evaluations. Hungarian Wikipedia conducted its own set of experiments by turning off FlaggedRevs, and then turning it back on after seeing no positive improvements in engagement metrics. This finding suggests that leaving implementation choices to each subcommunity might be preferable to reconcile technological frames across multiple subcommunities, in ways that make sense in Orlikowski and Gash's hierarchically managed context.

Next, interviewees described how their wikis tweaked FlaggedRevs extensively to meet their needs, solve technical and workload management issues, and align technological frames within their subcommunities.
Similar to Pending Changes on English Wikipedia, Farsi and Indonesian Wikipedias scaled back the scope of the system, resulting in an update that they are mostly happy with.
Although very different from Orlikowsi and Gash's \textit{Lotus Notes} context, this finding is in line with several studies of community-based moderation that point to the importance of user innovation in adapting to and improving moderation technology \cite{kiene2019technological, 10.1145/3338243}. 

\subsection{Implications}

Orlikowski and Gash's findings suggest the need for a bottom-up design process that takes into consideration the frames of managers and users. In accordance with Orlikowski and Gash's recommendations, FlaggedRevs went through multiple rounds of discussions where all community members could express their doubts or concerns. Despite this, FlaggedRevs struggled in many communities. 
Our results suggest that a grassroots approach does not guarantee the acceptance of a new tool. The personal nature of workload and workflow, and the sometimes-obtuse nature of technical implementation details, may be difficult to align with community expectations through discussion alone. The best solutions may involve giving communities the ability to experiment and tweak new technology on their own.

Our findings demonstrate implications for moderation decisions in a range of decentralized communities. 
For example, the rise of decentralized social media networks, such as Mastodon and Bluesky, raised questions about how these systems should be moderated. 
Flexible and decentralized forms of moderation may be less challenging to support in some ways because they do not require aligning technological frames across different communities into a single system.

Generally, our findings provide a framework that we believe can guide technological change in polycentric or self-governed communities. Our work suggests that communities seeking to deploy new technologies should conduct a careful review of new technology that considers the three dimensions we identified: platform challenges, issues that may arise among different stakeholders in communities, and technological challenges. The first two dimensions are unique to self-governed communities as they emerge from the governance structure and the communication necessary among the different layers. Further, the third is a natural consequence of technological change. Although it is far from a panacea, we believe that communities can use our framework to identify and potentially prevent problems arising from technological change.

\section{Conclusion}
\label{sec:conclusion}
By investigating the challenges in restructuring practices of content moderation on Wikipedia, our work highlights the delicate dynamics between disruptive technology and its stakeholders in a decentralized platform. These dynamics turned what could have been a success story into a cautionary tale. Although our work suggests that FlaggedRevs probably exhibits considerable flaws, it also underscores that a lack of attention to technological frames led to ineffective cooperation across levels of Wikipedia governance. Consequently, the deck was stacked against FlaggedRevs.
Even though many Wikipedia communities continue to see the usefulness of FlaggedRevs, their attempts to adapt and innovate with and around the system have been insufficient to address its limitations.

We hope that our work will help inform a more careful and socially informed approach to understanding context and the complexity of community-based moderation in the context of polycentric and self-governed communities. Moreover, the fact that FlaggedRevs continues to be successful in some language editions but not in others suggests that the system's problems are far from intractable. We hope that our study can enable more streamlined processes of communication between the platform and its members to align technological frames and support a successful rollout of similar systems in the future, in Wikipedia and beyond.

\section{Acknowledgements}

We owe a particular debt of gratitude to our interviewees who agreed to participate in our study and provided valuable insights. Our methodology was improved via generous feedback Andrea Forte, who has tremendous experience in conducting qualitative studies. The software used to analyze the manuscripts is hosted by the Community Data Science Collective, and the manuscript benefited from excellent feedback from several anonymous referees at ACM CSCW. This work was supported and funded by the National Science Foundation (grant number 2031951 and 2016061).

\bibliographystyle{ACM-Reference-Format}
\bibliography{main}

\received{July 2023}
\received[revised]{January 2024}
\received[accepted]{March 2024}
\end{document}